\documentclass{osa-article}

\journal{oe}
\articletype{Research Article}
\usepackage{lineno}
\usepackage{booktabs}

\begin{document}

\title{The state of optics research in Colombia: a scientometric analysis}

\author{Andres G. Marrugo,\authormark{1,*} Atilio Bustos-González,\authormark{2} and Egdar Rueda\authormark{3}}

\address{\authormark{1}Facultad de ingeniería, Universidad Tecnológica de Bolívar, Cartagena, Colombia\\
\authormark{2}SCImago Research Group, Medellín, Colombia\\
\authormark{3}Grupo de Óptica y Fotónica, Instituto de Física, Universidad de Antioquia, calle 70 No. 52 - 21, Medellín, Colombia}

\email{\authormark{*}agmarrugo@utb.edu.co} %% email address is required



\begin{abstract*}
Although there is a perception in the community that the state of optics research in Colombia has reached a mature state with international recognition, to date, there is no study that supports this view quantitatively. We aimed to assess the state of optics research in Colombia based on international journal publications. Therefore, we determined scientometric indicators using the research articles published by authors with Colombian affiliation in journals indexed in the Scopus database belonging to the Atomic and Molecular Physics and Optics subject category. The research output has increased dramatically in the past two decades, with an average of 169 articles per year since 2016. Most of these articles are published in high-impact journals. A little over 10\% of that research is in the top 10\% most cited in the world. Over 25 higher education institutions contribute significantly to this research with many international and national collaborations. The normalized citation impact for Colombian optics research is 0.95, only five points below the world average, and ranked second in Latin America, only superseded by Chile (1.33). Our results show that optics research is an established research area in Colombia with high impact and many active groups from different institutions spread throughout the country.
\end{abstract*}

%%%%%%%%%%%%%%%%%%%%%%%%%%  body  %%%%%%%%%%%%%%%%%%%%%%%%%%
\section{Introduction}
Optics is globally acknowledged as a scientific discipline that enriches other areas of science and technology. Since it is at the frontier of basic and applied science, it is often closely tied with innovations and technological impact~\cite{golnabi2006trend}. Optical technologies have been found to provide one of the most intensive interactions with science~\cite{ponds2009innovation}. In a way, assessing the state of optics research in a country is a form of probing the underlying research dynamics of a country. At present, there is only the perception that optics research in Colombia has reached a mature state with international recognition, but there are no studies that support this view quantitatively. 

In the last decade, scientometrics has emerged as a field of study concerned with measuring and analyzing the impact of scholarly literature for its use in policy and management contexts~\cite{leydesdorff2012scientometrics}. Most scientometric studies are often broad in terms of scientific field coverage~\cite{glanzel2018scientometric}, yet with the appropriate expert judgment, the analysis of individual fields can also be carried out~\cite{takeda2009optics}. In the field of optics, we found a study by Takeda and Kajikawa~\cite{takeda2009optics}, in which they built a citation network of papers to identify optics emerging research domains such as optical communication, quantum optics, optical data processing, optical analysis, and lasers. They also found the United States to be the leading country in publications and citations. Also, two recent studies by Kappi et al.,~\cite{kappi2020overview, kappi2020bibliometric} provided an overview of trends in Indian optics research from 2008 to 2019. Both studies found an exponential growth in the number of published articles, although the quality needs to improve. 

Regarding the Colombian scenario, many works have described the state of research~\cite{anduckia2000bibliometric, cortes2021colombian} from as early as the 1980s. However, it was not until the turn of the century that the Colombian research system started consolidating itself~\cite{de2008indicadores}. For the last 20 years, the Colombian Observatory of Science and Technology (OCyT) has been producing reports pointing in the same direction; notably, Colombian research output has doubled over the last decade~\cite{de2020indicadores}. However, field-specific analyses are necessary for accurately informing stakeholders and policy-makers. 

This work assesses the state of optics research in Colombia based on international journal publications. We determined scientometric indicators using the research articles published by authors with Colombian affiliation in journals indexed in the Scopus database belonging to the Atomic and Molecular Physics and Optics subject category. We provide an overall picture of the Colombian optics research landscape from the last two decades. However, we also concentrate on the 2016-2020 period to accurately estimate the current state. An important caveat of our results is that many researchers in optics often publish in other research fields, for example, in biomedicine or materials science; however, those articles are not included here. The following section provides a brief historical background of optics research in Colombia to connect better the current state with the underlying causes. In the subsequent sections, we show the results and extensively discuss the strengths and opportunities for improvement of Colombian optics research.

\section{Background}
This section presents a brief history of optics research and optics groups in Colombia from 1970 to 2002. Subsequently, we present a short analysis of the actual state and distribution of research groups with contributions to optics or related areas. To gather the information, we used the data available from the Colombian Optical Network Society (Sociedad Red Colombiana de Óptica) to contact the research groups directly and ask for information related to members, lines of research, and the group's history. We complemented the information provided by the research groups with the data available in GrupLac. Then, we searched in GrupLac all the research groups with names associated with optics or related areas such as photonics, spectroscopy, imaging, and lasers. Finally, using the database of authors with at least three publications in optics journals obtained with Scopus, we used CvLac and GrupLac to find the associated research groups for each author.   We built the groups' background, location, and establishment date with this information.

\begin{figure}[t]
\centering
\includegraphics[width=0.60\textwidth]{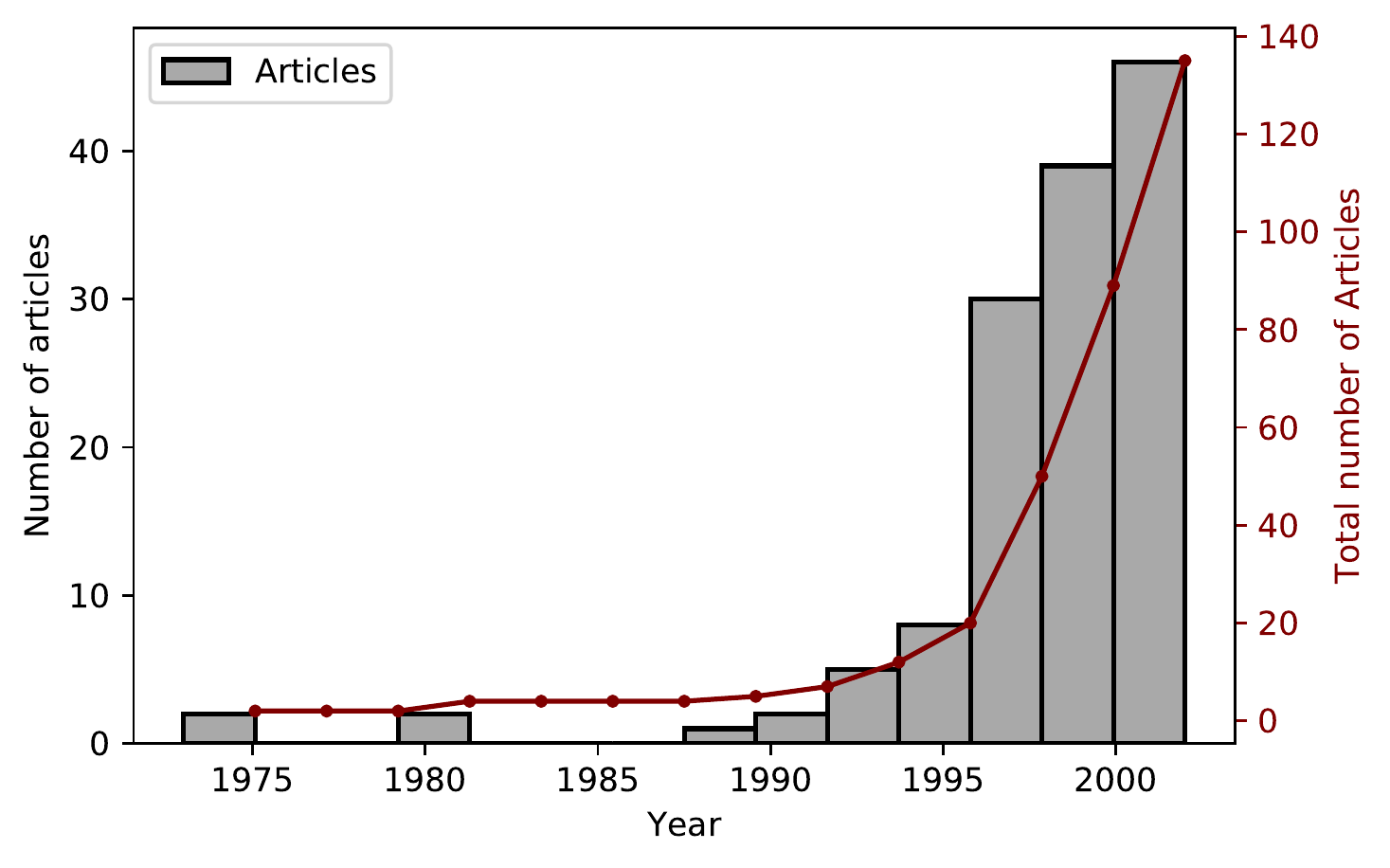}
\caption{The bars correspond to the number of articles published with affiliation to a Colombian Institution, while the line is used to show the total number of articles from 1973 to the specific year.}
\label{fig:ArtYearb2001}
\end{figure}

\subsection{History}
\begin{figure}[t]
\centering
\includegraphics[width=0.60\textwidth]{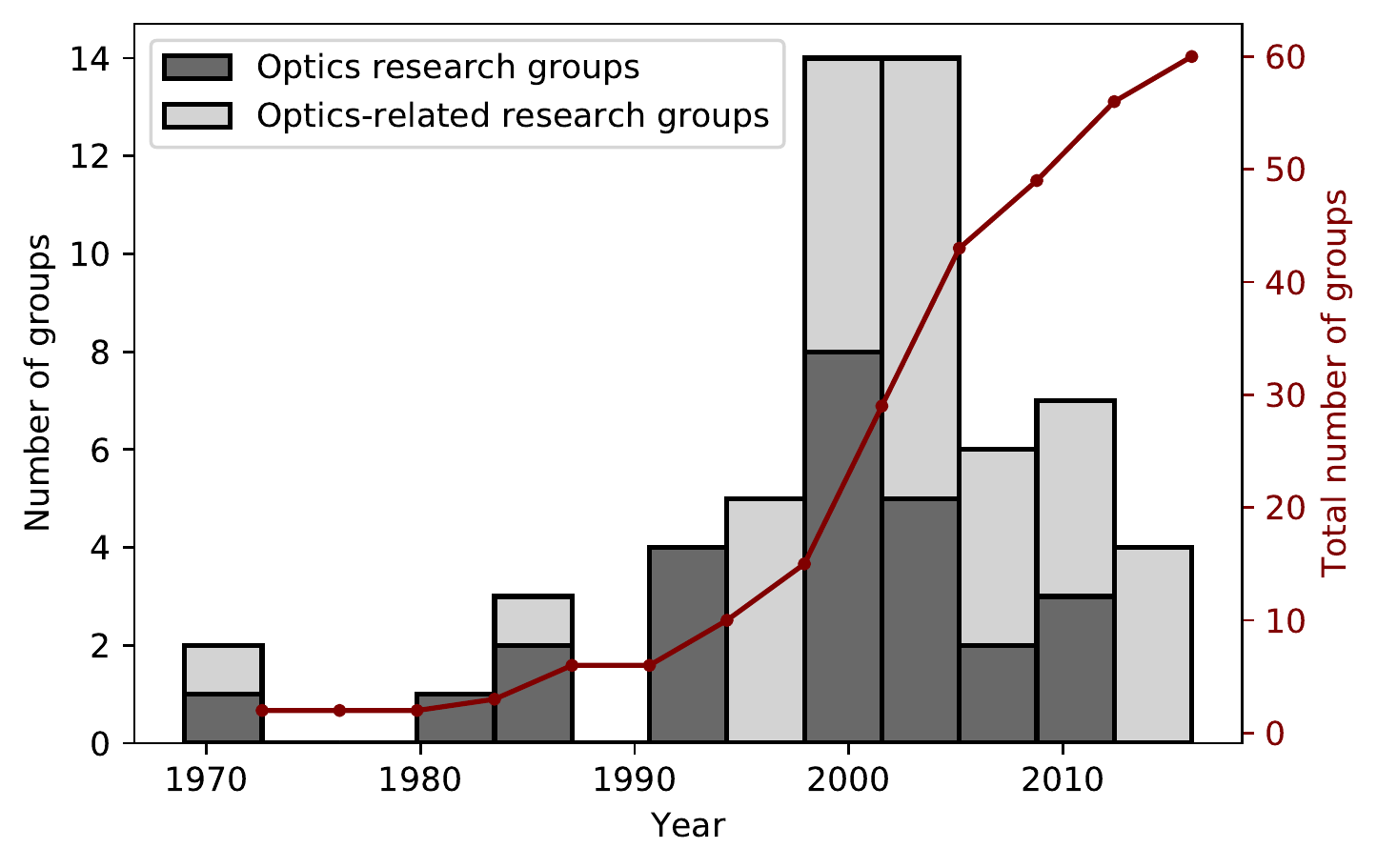}
\caption{The dark-grey bars correspond to the creation in time of research groups that consider themselves Optics research groups, while the light-grey bars correspond to research groups not directly associated to optics, but that have a research line in optics or related areas, or publications in AMPO. The line is used to show the increment of research groups associated with optics or related areas.}
\label{fig:gruposYear}
\end{figure}

The first physics departments in Colombia were created in the mid-1950s, and the 1960s~\cite{UdeAHist,UniAndes,UNalHist}, while the first known optics research groups appeared in the decades of 1970 and 1980. The oldest record we found of an optical research group corresponds to the Grupo de Óptica y Fotónica of the Universidad de Antioquia~\cite{GOF}. It was created by Austrian professor Dr. Peter Barlai in the year 1972. This group will later become the seed of many investigators, who, together with other Colombian researchers, will move forward the research in optics, including creating the Colombian Optical Network (Red Colombiana de Óptica). 
A decade had to pass for the appearance of new recognized optical groups: Espectroscopía Atómica y Molecular (1980) and Óptica y Tratamiento de Señales (1984) in Bucaramanga, and Óptica y Láser (1987) in Popayan. From this point, many new groups were created in the following decade until 2005, when the number of new optics groups started to decrease, as shown in Fig.~\ref{fig:gruposYear}). These timeframes coincide with the return to the country of many Colombians with a Ph.D. in Optics or related areas.  

The establishment of the research groups came with the first published scientific articles in optics with Colombian affiliations (Fig.~\ref{fig:ArtYearb2001}). The first records are from 1973 to 1981 with the work of Dr. Peter Barlai, who published in 1973 in the Journal of Physical Sciences: Zeitschrift für Naturforschung A~\cite{barlai1973koharenz},  the work of Dr. Alfonso Rueda, who joined in 1970 the Universidad de Los Andes~\cite{UniAndes} and published in 1973, and 1981, in Physical Review A~\cite{rueda1973fluctuations,rueda1981behavior}, and finally, Dr. Eddien Alvarez, who joined the Universidad de Antioquia in 1967, obtained his Ph.D. in 1980~\cite{UdeA}, and published in 1980 in Physical Review A~\cite{alvarez1980hyperfine}.

After eight years without publications, from 1989 onward, the number of published articles started to climb rapidly, reaching a total of 135 optics research articles by the end of the year 2002, as shown in Fig.~\ref{fig:ArtYearb2001}.
As we explained earlier, the return of many researchers with a Ph.D. in optics could explain the boost in productivity alongside newly created funding policies from institutions and the government. Notwithstanding the productivity increase, the establishment of new groups stalled in the last decade. This situation might have two explanations. New optics researchers were being incorporated into already established groups, or the funding policies favored established groups with long-standing publication records. Either way, the conditions for establishing new groups by young principal investigators are not what they used to be a decade ago.

In parallel to the establishment of research groups in optics, from 1970 to 2020, other groups were established in optics-related fields, despite not defining themselves as optics research groups. Typically, these groups have only a single research line in optics or related fields, or some of their publications appear in optics journals (see Fig. \ref{fig:gruposYear}). These groups are mainly associated with engineering, condensed matter, applied physics, electronics, materials science, and instrumentation. 

\subsection{Red SCienTi analysis} 
\begin{figure}[!htb]
\centering
\includegraphics[width=0.60\textwidth]{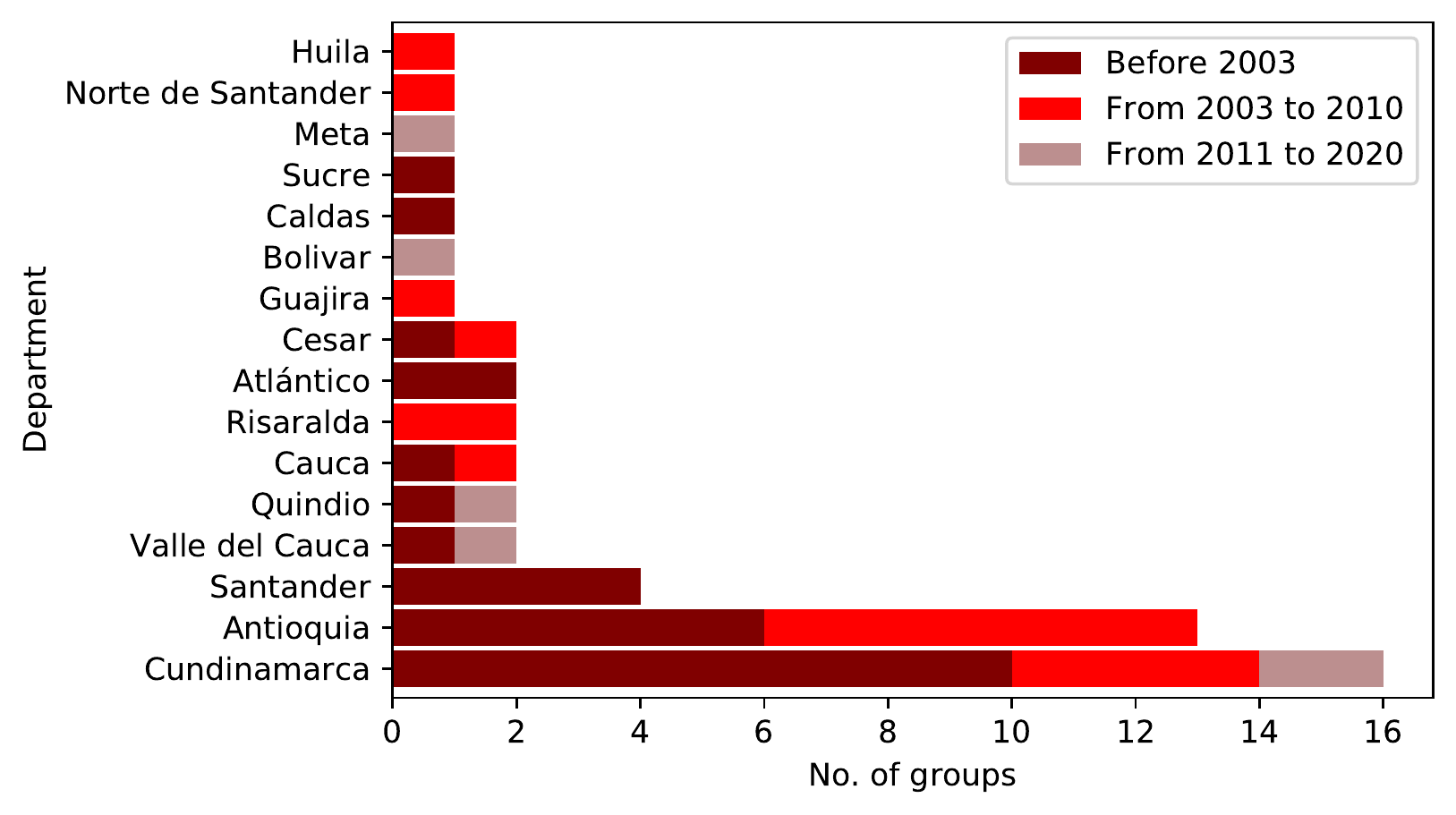}
\caption{Distribution in Colombia of research groups with at least one research field in optics or  optics-related field.}
\label{fig:gruposLoc}
\end{figure}

ScienTI is a public network of academic and research information that contributes to the management of scientific and technological development of its members~\cite{SCienTi}. The network includes information about researchers (CvLAC), research groups (GrupLAC), and Institutions (InstituLAC). In Colombia, the Ministry of Science, Technology, and Innovation (Minciencias) is in charge of the Colombian branch of the network through the Sistema Nacional de Ciencia, Tecnología, e Innovación (SNCTI)~\cite{abcMin}. In particular, Minciencias has a program to identify and measure the research groups in the Colombian Science, Technology and Innovation Network to create and analyze funding policies and strategies in order to generate knowledge and technology that will impact the development and growth of the country positively~\cite{GrupLAC}. Of the 5772 research groups recognized by SNCTI in 2020, we have found 61 research groups identified as optics groups or with at least one researcher with three or more publications in optics journals between 2016-2020. From the 61, 26 were identified as optical groups, and 52 have at least one research line in optics or related fields. The majority of these groups are concentrated in the central part of the Colombian Andes, likely for historical reasons, with a few distributed in the northern part of the country (see Fig.~\ref{fig:gruposLoc}). The top three administrative departments with the most optics or optics-related research groups are Cundinamarca, Antioquia, and Santander.

\section{Methods}
\label{sec:methods}

\iffalse
Our analyses are based on two approaches. First, we used the data available from the RedScienti platform~\cite{scienti} from the  Colombian Ministry of Science, and the data available from the Colombian Optical Society (Sociedad Red Colombiana de Óptica)~\cite{rco}, to get an overview of the existing optics research groups in the country. Second, we used the Scopus database to carry out a scientometric analysis from Colombian authors and institutions.
\subsection{Scientometric analysis}
\fi

For all of our analysis, we used the Scopus database. We considered two analysis levels to provide different scale-dependent perspectives on the state of optics research in Colombia:

\paragraph{Category-level analysis.} 
This analysis consists of all publications in the Scopus journal category ``Atomic and Molecular Physics, and Optics'' (AMPO) with at least one author with Colombian affiliation from the year 2003 to 2020. This journal category belongs to the larger subject area ``Physics and Astronomy''. This classification allows to evaluate and compare the impact of research of countries or institutions within the same scientific field. This procedure allowed us to analyze the impact of Colombian researchers, to compare it against  their peers from other countries, and determine overall trends in the past two decades. To better assess the current state of optics research in Colombia, we focused on the 2016-2020 period. We determined several scientometric indicators, we identified the top contributing Colombian institutions, and described the potential for technological impact by identifying the top cited articles by patents.  

\paragraph{Selected journals analysis.} 
Although the above approach may serve to get an overall picture of the Colombian research for the AMPO category, it has several shortcomings as not all journals in the AMPO category are strictly optics journals. Using expert judgment~\cite{wang2016large}, we carried out a selection of 131 journals from the 239 journals in the AMPO category related exclusively to optics (the journal selection data is available in the external repository). 
We determined the top contributing institutions and the active researchers (defined as authors with 3 or more articles within 2016-2020) in each institution with corresponding scientometric indicators. This selection-based approach allowed us to get a more detailed picture of which institutions have the most active researchers, which of them achieve higher impacts or depend more heavily on international collaboration, among other relevant aspects. Moreover, we also determined the journals with most of the articles between 2016-2020.

\subsection{Scientometric indicators}

The scientometric indicators used in our study were the following.

\begin{itemize}
    \item \textbf{Output:} The total number of documents published in scholarly journals indexed in Scopus~\cite{bornmann2008citation, romo2011analysis, oecd2016compendium}. Size-dependent indicator.
    % (Bornmann , L.oed, Mutz , Rudiger., Neuhaus , Christoph., & Daniel, Hans., 2008; Romo-Fernández, Luz; López-Pujalte, Cristina; Guerrero-Bote, Vicente; Moya-Anegón, Félix, 2011; OECD, 2016).
    \item \textbf{Scientific Leadership:} Leadership indicates the amount of an institution’s output as main contributor, that is, the amount of papers in which the corresponding author belongs to the institution~\cite{moya2012liderazgo, moya2013research}. Size-dependent indicator.
    % (Moya-Anegón, 2012; Moya-Anegón, F., Guerrero-Bote, V., Bornmann, L., Moed, H.F, 2013).
    \item \textbf{International Collaboration:} Institution's output produced in collaboration with foreign institutions. The values are computed by analyzing an institution's output whose affiliations include more than one country address~\cite{chinchilla2010new, lancho2012citation, guerrero2013quantifying}. Size-dependent indicator. 
    % (Chinchilla-Rodríguez, Z., B. Vargas-Quesada, Y. Hassan-Montero, A. González-Molina, and F. Moya-Anegón., 2010; Lancho-Barrantes, B. S., Guerrero-Bote, V. P., Chinchilla-Rodríguez, Z., Moya-Anegón, F., 2012; Guerrero-Bote, Olmeda-Gómez and Moya- Anegón, 2013).
    \item \textbf{High Quality Publications (Q1):} the number of publications that an institution publishes in the most influential scholarly journals of the world. These are those ranked in the first quartile (25\%) in their categories as ordered by SCImago Journal Rank (SJRII) indicator~\cite{miguel2011open, chinchilla2016benchmarking}. Size-dependent indicator.
    % (Miguel, Chinchilla-Rodríguez and Moya-Anegón, 2011; Chinchilla-Rodríguez, Miguel, and Moya-Anegón, 2016). 
    \item \textbf{Normalized Impact (NI):} Normalized Impact is computed over the institution's output using the methodology established by the Karolinska Institutet in Sweden where it is named "Item oriented field normalized citation score average". The normalization of the citation values is done on an individual article level. The values (in decimal numbers) show the relationship between an institution's average scientific impact and the world average set to a score of 1, --i.e. a NI score of 0.8 means the institution is cited 20\% below world average and 1.3 means the institution is cited 30\% above average~\cite{rehn2008bibliometrics, gonzalez2010new, guerrero2012further}. Size-independent indicator. 
    % (Rehn and Kronman, 2008; González-Pereira, Guerrero-Bote and Moya-Anegón, 2010; Guerrero-Bote and Moya-Anegón, 2012).
    \item \textbf{Excellence10:} Excellence indicates the amount of an institution’s scientific output that is included in the top 10\% of the most cited papers in their respective scientific fields. It is a measure of high quality output of research institutions~\cite{bornmann2012new, bornmann2014proportion, bornmann2014ranking}. Size-dependent indicator.
    % (Bornmann, Moya-Anegón and Leydesdorff, 2012; Bornmann and Moya-Anegón, 2014a; Bornmann, L., Stefaner, M., Moya Anegón, F., &Mutz, R., 2014b). 
    \item \textbf{Excellence10 with Leadership:} Excellence with Leadership indicates the amount of documents in Excellence in which the institution is the main contributor~\cite{bornmann2012new}. Size-dependent indicator. 
    % (Bornmann, L., De Moya Anegón, F., Leydesdorff, L., 2012). 

\end{itemize}

To make the above indicators size-independent, it is customary to express them as a percentage of the total output, for example, when referring to ``\% International Collaboration'' we mean the percentage of articles with international collaboration relative to the total output of the institution.
% \% Excellence10 we mean the percentage of articles relative to the total output that are in the top 10\% most cited in the world. 

\subsection{Data availability}
All data underlying the findings are fully available without restriction. The final datasets and accompanying code are available on the Open Science Framework DOI \url{https://osf.io/hecgw/?view_only=3f78bd27bf28476c968da9cf0582a871}.

\section{Scientiometric analysis results}

\subsection{Category-level analysis.}

\paragraph{The big picture.}

\begin{figure}[b]
\centering
\includegraphics[width=0.9\textwidth]{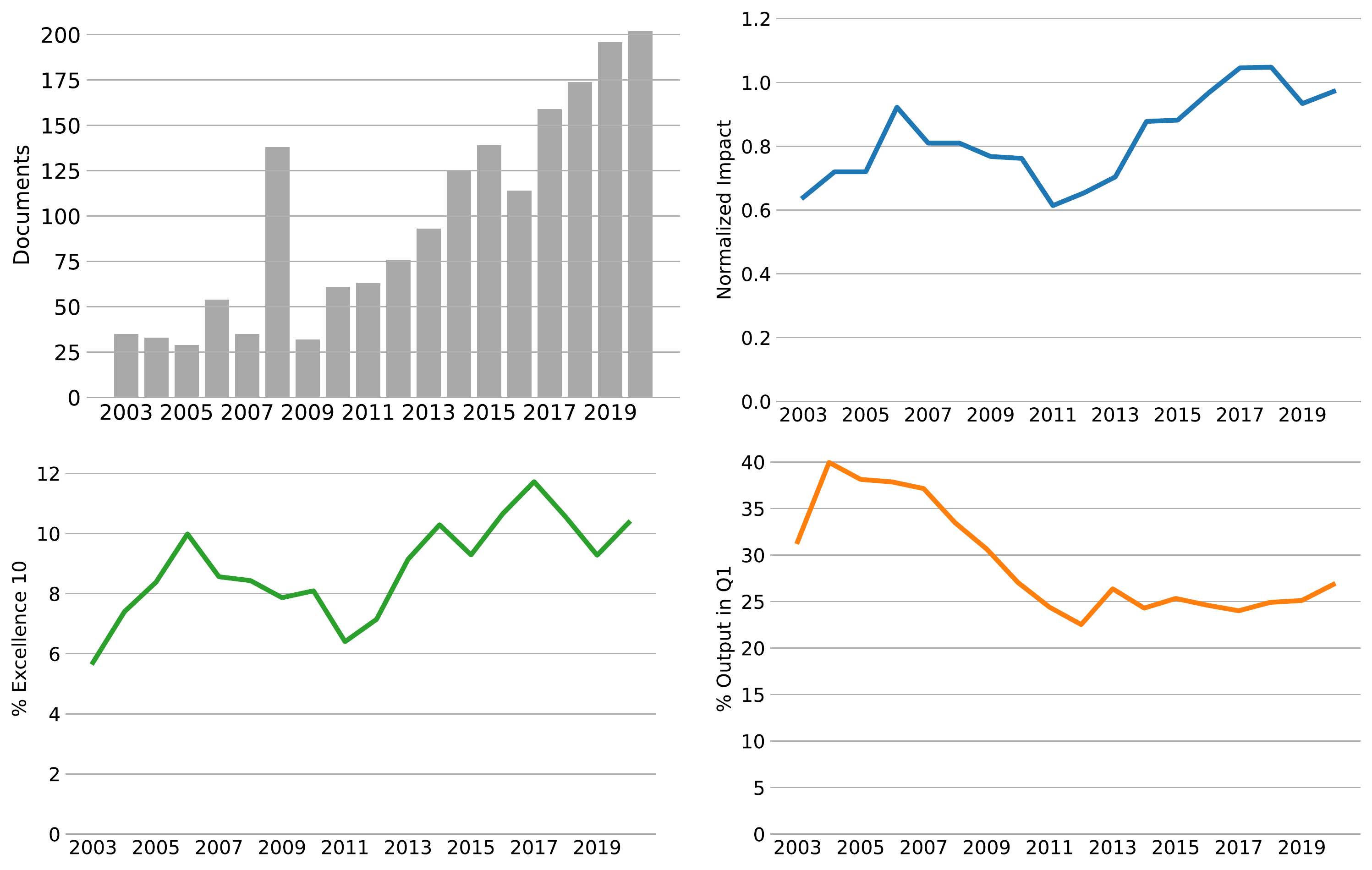}
\caption{Number of published journal articles in AMPO Colombia between 2003 and 2020, and selected scientometric indicators (5-year moving average).}
\label{fig:selected-indicators}
\end{figure}

Let us begin by providing an overview of how the optics research output from Colombia has changed in the past two decades. In Fig.~\ref{fig:selected-indicators}, we show the number of published articles in AMPO journals with at least one author with a Colombian affiliation from 2003 to 2020. The research output has been steadily rising, with over 100 articles per year since 2014 and a little over 200 in 2020. These numbers are in sharp contrast with the 135 articles published by all Colombian authors in optics journals from 1973 to 2000. 
% \xout{In a way, it could be said that Colombian research in optics took off in the mid-2000s}.

In Fig.~\ref{fig:selected-indicators}, we also plot several scientometric indicators to show the 5-year moving average tendency for the same period. The normalized citation impact has been increasing along with the research output, surpassing the world average impact in AMPO between 2017 and 2018. We observe the same tendency for the Excellence10 indicator, between 10\% and 12\% from 2015 to 2020. Remarkably, the normalized impact or the excellence10 indicators have not declined with the increase in output; instead, they have risen.

The same tendency has not happened for the percentage of articles in Q1 journals. Although the output in the 2000s was relatively low, a significant proportion of those articles were published in Q1 journals (30\% to 40\%). From 2012 to 2020, this percentage has stabilized around 25\%. 

\paragraph{The current state.}

Now, let us focus on the current state of the research by providing the indicators between 2016 to 2020. These are shown in table~\ref{tab:indicators}. The full 2003 to 2020 data are included in table~\ref{tab:01a} as an appendix. From 2016 to 2020, 845 articles were published with 5471 citations, accounting for an average of 169 articles per year and 6.5 citations per article.

Other relevant aspects are the relatively high percentage of cited articles of 68.6 \%, the international collaboration of 61.1\%, the percentage of output in Q1 of 26.5\%, and the percentage of leadership articles of 68.8\%. This last indicator suggests that Colombian optics researchers have achieved significant independence from their international collaborators. 

The normalized citation impact was 0.95, indicating that the articles from Colombian researchers were cited only 5\% below the world average. When compared to the AMPO research output from other similar countries in Latin America with which Colombian researchers collaborate, we found that Colombia ranks second behind Chile (1.33), followed by Brazil (0.88), Argentina (0.80), and Mexico (0.77)~\cite{scimago_2021}. The excellence10 indicator also reveals the quality of work being conducted in the country with a 10.1\%. Unfortunately, this indicator falls to 4.5\% when considering Colombian leadership. This fact hints that international collaboration is partly responsible for highly cited articles.

\begin{figure}[!h]
\centering
\includegraphics[width=0.55\textwidth]{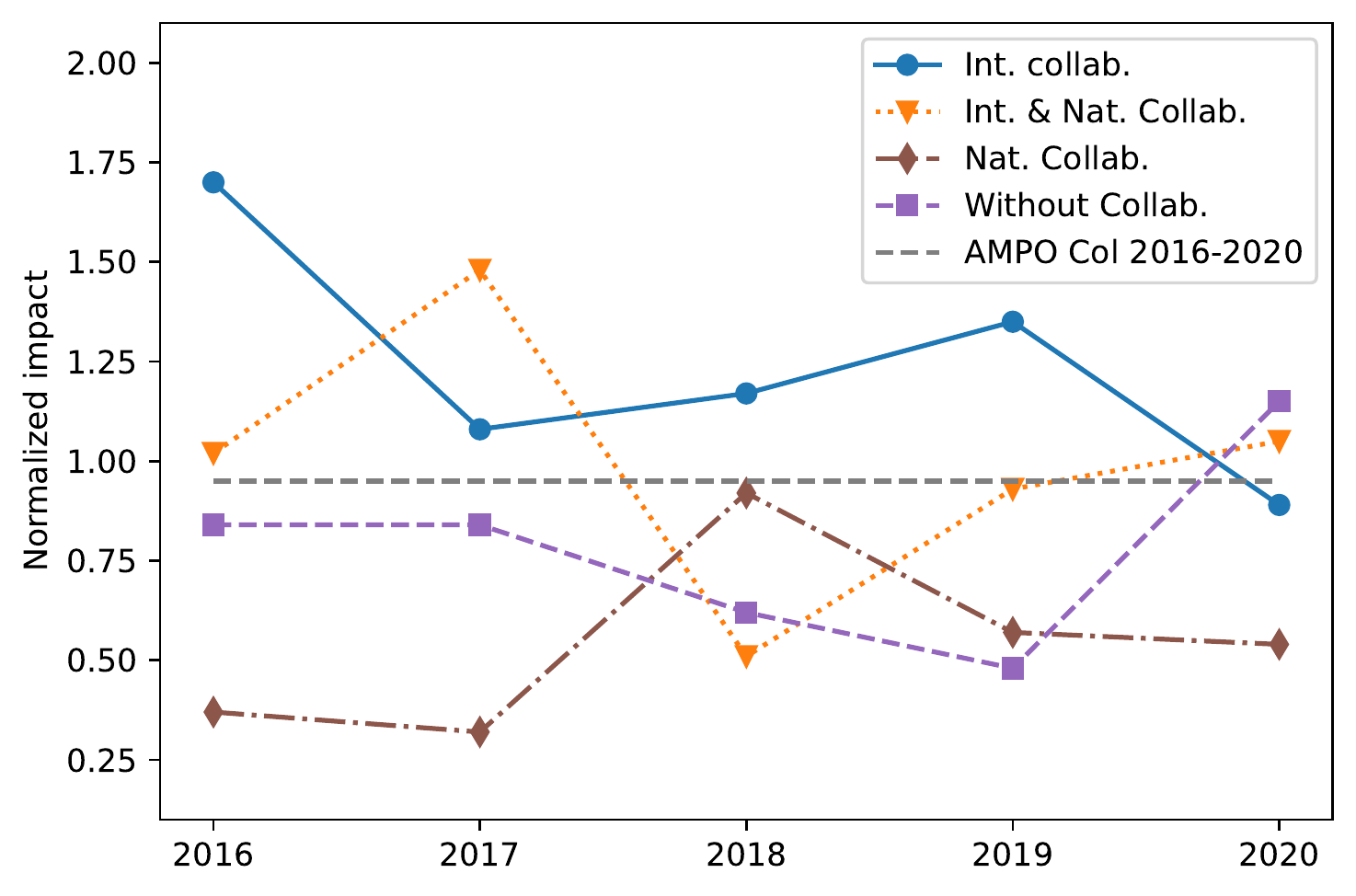}
\caption{Normalized impact related to the type of collaboration from 2016-2020. The normalized impact for AMPO Colombia is 0.95, however, articles published with international collaboration only and international and national collaboration are the ones that consistently achieve normalized impact above AMPO Colombia and the world.}
\label{fig:collab}
\end{figure}

Although the normalized impact of the whole country is 0.95 between 2016-2020, we can show how the normalized impact varied year after year depending on the type of collaboration. In Fig.~\ref{fig:collab}, we show that articles published with international collaboration-only and international-and-national collaboration are the ones that most consistently achieve a normalized citation impact above AMPO Colombia and the world average. No collaboration or strictly national collaboration more often achieved below AMPO Colombia normalized citation impact.

Articles published in open access journals have become an essential aspect in assessing the state of research. The percentage output in open access journals was 27.1\%, which is somewhat below the 50.3\%  for all subject categories for Colombia~\cite{scimago_2021}.

\begin{table}[t]
\centering
\begin{tabular}{@{}lllllll@{}}
\toprule
Indicator                      & 2016 & 2017 & 2018 & 2019 & 2020 & Total \\ \midrule
Output                         & 114  & 159  & 174  & 196  & 202  & 845   \\
Cites                          & 1734 & 1628 & 1227 & 700  & 182  & 5471  \\
Cites per document             & 15.2 & 10.2 & 7.1  & 3.6  & 0.9  & 6.5   \\
\% Cited documents             & 91.2 & 80.5 & 82.8 & 63.8 & 39.1 & 68.6  \\
\% International collaboration & 64.9 & 67.9 & 62.6 & 51.0 & 61.9 & 61.1  \\
Output in Q1                   & 23   & 35   & 64   & 49   & 61   & 232   \\
\% Output in Q1                & 20.2 & 22.0 & 36.8 & 25.0 & 30.2 & 26.5  \\
Leadership                     & 77   & 111  & 118  & 133  & 142  & 581   \\
\% Leadership                  & 67.5 & 69.8 & 67.8 & 67.9 & 70.3 & 68.8  \\
Normalized Impact              & 1.22 & 0.95 & 0.90 & 0.89 & 0.90 & 0.95  \\
Normalized Impact wL           & 0.71 & 0.86 & 0.72 & 0.56 & 0.78 & 0.72  \\
Excellence10                   & 15   & 19   & 16   & 11   & 24   & 85    \\
\% Excellence10                & 13.2 & 12.0 & 9.2  & 5.6  & 11.9 & 10.1  \\
Excellence10 wL                & 5    & 9    & 5    & 6    & 13   & 38    \\
\% Excellence10 wL             & 4.4  & 5.7  & 2.9  & 3.1  & 6.4  & 4.5   \\
Open Access                    & 24   & 48   & 53   & 47   & 57   & 229   \\
\% Open Access                 & 21.1 & 30.2 & 30.5 & 24.0 & 28.2 & 27.1  \\ \bottomrule
\end{tabular}
\caption{General productivity and scientometric indicators for the Atomic and Molecular Physics, and Optics (AMPO) category for Colombia between 2016 and 2020.}
\label{tab:indicators}
\end{table}

% \begin{figure}[!ht]
% \centering
% \includegraphics[width=0.60\textwidth]{figures/fig-docs-ex10.pdf}
% \caption{Number of published journal articles (bars) and percentage in top 10\% most cited in the world between 2003-2020 (5-year moving average) (red line). In the past 5 years, the \% Excellence10 indicator has stabilized between 10-12\%.}
% \label{fig:docs-ex10}
% \end{figure}

\paragraph{Institutions.}

In table~\ref{tab:ampo_institutions}, we shot the top 25 contributing institutions in AMPO Colombia between 2016-2020. They are all higher education institutions. In the first place, we find Universidad Nacional de Colombia (UNAL) with 538 articles. Note that UNAL is the largest public university in Colombia, with several campuses throughout the territory, and in this analysis they all appear unified as a single institution. The top 5 is completed by Universidad de Antioquia (208), Universidad de los Andes (85), Universidad Industrial de Santander (82), and Universidad del Valle (82). However, if we were to rank the institutions by citations per document, the top 5 institutions were Universidad Antonio Nariño (26.7), Universidad Tecnológica de Pereira (20.6), Universidad EIA (14.0), Universidad de Antioquia (13.13), and Universidad de los Andes (9.36). Notably, in raw numbers, the Universidad de Antioquia stands out. 

Nevertheless, it is also remarkable that many universities spread throughout the territory have noteworthy contributions. This aspect reveals the potential for establishing stronger national alliances, inter-institution exchange programs, and other initiatives.

\begin{table}[!ht]
\begin{tabular}{@{}lllll@{}}
\toprule
No & Institution                                      & Documents & Cites & Cites per document \\ \midrule
1  & Universidad Nacional de Colombia                  & 538       & 3497  & 6.50               \\
2  & Universidad de Antioquia                          & 208       & 2730  & 13.13              \\
3  & Universidad de los Andes                          & 85        & 796   & 9.36               \\
4  & Universidad Industrial de Santander               & 82        & 647   & 7.89               \\
5  & Universidad del Valle                             & 82        & 482   & 5.88               \\
6  & Instituto Tecnologico Metropolitano               & 61        & 414   & 6.79               \\
7  & Universidad Pontificia Bolivariana                & 39        & 231   & 5.92               \\
8  & Universidad Surcolombiana                        & 38        & 130   & 3.42               \\
9  & Universidad EAFIT                                & 38        & 115   & 3.03               \\
10 & Universidad Pedagogica y Tecnologica de Colombia & 34        & 139   & 4.09               \\
11 & Universidad Tecnologica de Pereira                & 33        & 679   & 20.58              \\
12 & Universidad Tecnologica de Bolivar               & 22        & 109   & 4.95               \\
13 & Universidad EIA                                  & 21        & 294   & 14.00              \\
14 & Universidad del Atlantico                        & 20        & 122   & 6.10               \\
15 & Universidad de Medellin                          & 20        & 42    & 2.10               \\
16 & Politecnico Colombiano Jaime Isaza Cadavid        & 19        & 97    & 5.11               \\
17 & Universidad Popular del Cesar                    & 18        & 86    & 4.78               \\
18 & Universidad del Quindio                           & 18        & 75    & 4.17               \\
19 & Universidad del Magdalena                        & 17        & 62    & 3.65               \\
20 & Universidad Autonoma de Occidente                & 16        & 67    & 4.19               \\
21 & Universidad del Norte                            & 16        & 117   & 7.31               \\
22 & Universidad Cooperativa de Colombia              & 14        & 136   & 9.71               \\
23 & Universidad de la Costa                          & 13        & 50    & 3.85               \\
24 & Universidad Antonio Nariño                       & 13        & 348   & 26.77              \\
25 & Pontificia Universidad Javeriana                 & 13        & 85    & 6.54               \\ \bottomrule
\end{tabular}
\caption{Colombian institutions with more than 10 documents in AMPO between 2016-2020. }
\label{tab:ampo_institutions}
\end{table}

\paragraph{Potential for technological impact}

Often, an overlooked aspect in many studies is the potential for technological impact. However, because Optics is a field in the frontier between basic and applied science, we consider the potential technological impact of articles from AMPO Colombia in terms of citations from patents. In table~\ref{tab:cited_patents} we show the 15 documents from Colombian researchers with two or more citations from patents. Note that even conference papers (Proc. SPIE) have been highly cited, revealing that potential technological impact is less related to conventional article metrics.  We have analyzed a larger window from 2011 to 2020, considering that these citations take more time to accumulate than regular citations from other journal articles.

\begin{table}[!ht]
\centering
\begin{tabular}{@{}lp{7.5cm}lp{2.5cm}l@{}}
\toprule
No & Title                                                                                                                                     & Year & Source                               & Patents \\ \midrule
1  & Optical encryption and QR codes: Secure and noise-free information retrieval                                                              & 2013 & Optics Express                       & 7       \\
2  & Automatic detection of invasive ductal carcinoma in whole slide images with convolutional neural networks                                 & 2014 & Proc. SPIE                           & 6       \\
3  & Distributed Dynamic Host Configuration Protocol (D2HCP)                                                                                   & 2011 & Sensors                              & 6       \\
4  & Multi-view information fusion for automatic BI-RADS description of mammographic masses                                                    & 2011 & Proc. SPIE                           & 5       \\
5  & Color lensless digital holographic microscopy with micrometer resolution                                                                  & 2012 & Optics Letters                       & 5       \\
6  & Higher-order computational model for coded aperture spectral imaging                                                                      & 2013 & Applied Optics                       & 4       \\
7  & Multi-wavelength digital in-line holographic microscopy                                                                                   & 2012 & OSA DTu1C. 4                         & 4       \\
8  & Cascaded ensemble of convolutional neural networks and handcrafted features for mitosis detection                                         & 2014 & Proc. SPIE                           & 4       \\
9  & Compressive spectral polarization imaging by a pixelized polarizer and colored patterned detector                                         & 2015 & JOSA A                               & 3       \\
10 & High-Q silicon nitride microresonators exhibiting low-power frequency comb initiation                                                     & 2016 & Optica                               & 3       \\
11 & Structural and hyperfine properties of Mn and Co-incorporated akaganeites                                                                 & 2014 & Hyperfine Interactions               & 2       \\
12 & Development of pillared clays for wet hydrogen peroxide oxidation of phenol and its application in the posttreatment of coffee wastewater & 2012 & Int. J. of Photoenergy & 2       \\
13 & Beam selection in multiuser millimeter-wave systems with sub-array user terminal architectures                                            & 2017 & IMOC 2017                            & 2       \\
14 & Shift-variant digital holographic microscopy: Inaccuracies in quantitative phase imaging                                                  & 2013 & Optics Letters                       & 2       \\
15 & Delay/disruption tolerant network-based message forwarding for a river pollution monitoring wireless sensor network application           & 2016 & Sensors                              & 2       \\ \bottomrule
\end{tabular}
\caption{Top cited documents in patents from AMPO Colombia between 2011-2020.}
\label{tab:cited_patents}
\end{table}

\subsection{Selected journals analysis.}
Even though journal classification systems facilitate bibliometric studies, they often fail to provide a refined view of a research field. The selection of optics-exclusive journals allowed us to obtain valuable data on the current state of optics research in Colombia. From the selection, we obtained 499 journal articles from 2016-2020 (482 articles, 15 reviews, 1 erratum, and 1 letter). The majority were written in English (495) and only a few in Spanish (4). 

\begin{figure}[!t]
\centering
\includegraphics[width=0.70\textwidth]{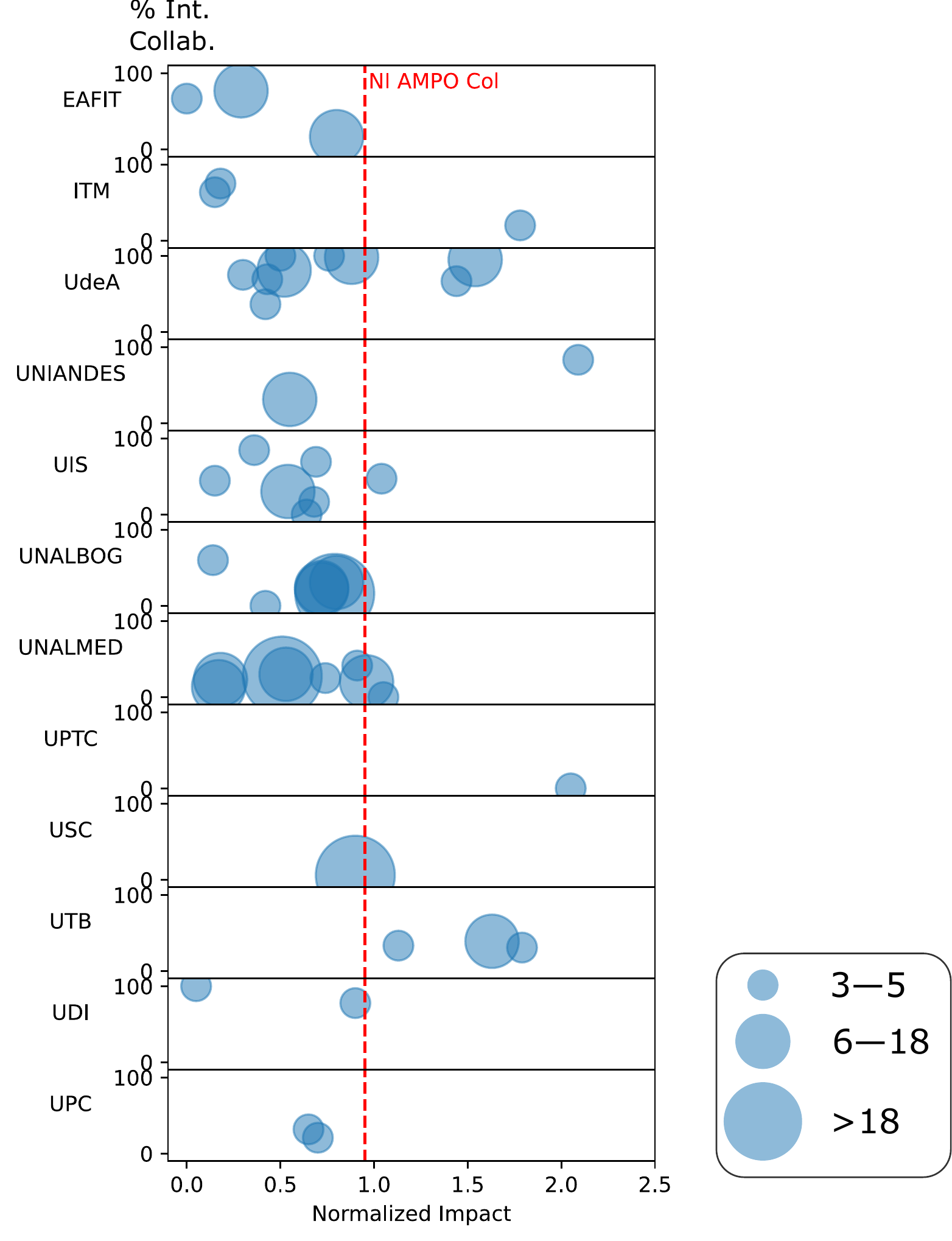}
\caption{Authors with at least three articles between 2016-2020 with their corresponding percentage International collaboration versus normalized citation impact. }
\label{fig:int-col-vs-ni}
\end{figure}

\paragraph{Authors}
The articles were written on average by 5.2 authors (min 2, max 29), showing that single-author papers are uncommon in the field. In Fig.~\ref{fig:int-col-vs-ni}, we show a summary of the active Colombian researchers, defined as authors with three or more articles between 2016-2020. They are grouped by affiliation as stacked plots. Each author is a circle, with its radius proportional to the research output. The horizontal axis is the normalized citation impact, and the vertical axis is the percentage of international collaboration. Plotted as a vertical red dashed vertical line is the normalized impact for AMPO Colombia. Some institutions appear in table~\ref{tab:ampo_institutions} but not here because their authors did not meet the threshold.

For this analysis, we were able to identify authors from each UNAL Campus individually. Note that there are institutions with a large base of optics researchers, such as UNAL Bogotá (UNALBOG), UNAL Medellín (UNALMED), or Universidad Industrial de Santander (UIS). However, their impact is often below 1.0 or below AMPO Colombia (0.95), regardless of the research output (circle size) or international collaboration (vertical axis). In sharp contrast, we find authors from institutions with fewer researchers, but with high impact such as those from Universidad Tecnológica de Bolívar (UTB), Universidad Pedagógica y Tecnológica de Colombia (UPTC), or Universidad de los Andes (UNIANDES). We also find a varied dependency on international collaboration. Universidad de Antioquia (UdeA) is somewhat atypical in that it has two clusters, one of the authors with impact below 1.0 and the other with an impact higher than 1.5. All UdeA authors have significant international collaboration, with some authors at 100\%. Instituto Tecnológico Metropolitano (ITM) has a similar situation, albeit with fewer authors. The other remarkable situation occurs with Universidad Surcolombiana, with a single highly prolific author with an impact close to 1. There are other notable mentions, such as Universidad Popular del Cesar (UPC), Universidad de Investigación (UDI), and EAFIT.

% \begin{figure}[!ht]
% \centering
% \includegraphics[width=\textwidth]{figures/fig-tab-top-journals-.pdf}
% \caption{Top 20 journals with most published articles 2016-2020. }
% \label{fig:top-journals}
% \end{figure}

\paragraph{Journals}
In table~\ref{tab:ampo_journals}, we show the top 30 journals with the most published articles between 2016-2020. The journal at the top of the list is Sensors with 104 articles, followed by Optik (63), Physical Review A (51), Applied Optics (38), and Optica Pura y Aplicada (29) to complete the top 5. Note that all journals in the top 30 list are high-impact journals (Q1 or Q2), except for two Q3 and Q4 journals. It is noteworthy that Optica appears in this list, even with only three documents, knowing that it aims to publish high-profile research in both fundamental and applied optics and photonics. It also has the highest Source Normalized Impact per Paper (SNIP) of 3.333.

\begin{table}[!ht]
\begin{tabular}{@{}lp{7cm}lcl@{}}
\toprule
No & Journal                                                    & SNIP                      & SJR Quartile         & No. Documents \\ \midrule
1  & Sensors Switzerland                                        & 1.555                     & Q2                   & 104           \\
2  & Optik                                                      & 0.875                     & Q2                   & 63            \\
3  & Physical Review A                                          & 0.993                     & Q1                   & 51            \\
4  & Applied Optics                                             & 1.038                     & Q2                   & 38            \\
5  & Optica Pura y Aplicada                                     & 0.178                     & Q4                   & 29            \\
6  & Optics Letters                                             & 1.432                     & Q1                   & 21            \\
7  & Optics Communications                                      & 0.918                     & Q2                   & 16            \\
8  & Optics Express                                             & 1.541                     & Q1                   & 15            \\
9  & Optical Engineering                                        & 0.691                     & Q3                   & 14            \\
10 & Optics and Lasers in Engineering                           & 2.328                     & Q1                   & 13            \\
11 & Journal of Optics United Kingdom                           & 0.967                     & Q1                   & 12            \\
12 & Journal of the Optical Society of America A                & 1.082                     & Q2                   & 12            \\
13 & Optical Materials                                          & 0.955                     & Q2                   & 12            \\
14 & Journal of the Optical Society of America B                & 0.842                     & Q2                   & 8             \\
15 & Photonics                                                  & 1.084                     & Q2                   & 8             \\
17 & Journal of Lightwave Technology                            & 1.778                     & Q1                   & 7             \\
18 & Journal of Luminescence                                    & 0.945                     & Q2                   & 7             \\
19 & Photonics and Nanostructures Fundamentals and Applications & 1.010                     & Q2                   & 6             \\
20 & Journal of Physics B Atomic Molecular and Optical Physics  & 0.819                     & Q2                   & 5             \\
21 & European Physical Journal D                                & 0.660                     & Q3                   & 4             \\
22 & International Journal of Photoenergy                       & 0.849                     & Q3                   & 4             \\
23 & Journal of Nonlinear Optical Physics and Materials         & 0.630                     & Q4                   & 4             \\
24 & Optics and Laser Technology                                & 1.427                     & Q2                   & 4             \\
25 & Biomedical Optics Express                                  & 1.550                     & Q1                   & 3             \\
26 & IEEE Photonics Technology Letters                          & \multicolumn{1}{r}{1.008} & Q1                   & 3             \\
27 & IEEE Transactions on Electromagnetic Compatibility         & 1.683                     & Q2                   & 3             \\
28 & ISPRS Journal of Photogrammetry and Remote Sensing         & 3.054                     & Q1                   & 3             \\
29 & Optica                                                     & 3.333                     & Q1                   & 3             \\
30 & IEEE Photonics Journal                                     & 0.997                     & Q2                   & 2             \\
   &                                                            &                           & \multicolumn{1}{l}{} &               \\ \bottomrule
\end{tabular}
\caption{Top 30 journals with most published articles between 2016-2020. SNIP: Source Normalized Impact per Paper.}
\label{tab:ampo_journals}
\end{table}

\section{Discussion}

%\subsection{The big picture}
% \subsection{Category-level analysis}

% \paragraph{The big picture.}

As we have shown, the number of published articles has changed dramatically throughout different periods but correlates well with the number of research groups. From 1970 up until 1990, there were but a handful of articles published by the first research groups in the country. From 1990 up until the mid-2000s, research productivity plateaued--even while new research groups were being created--probably because most researchers were university professors with hardly any graduate students at the time when most graduate programs were in their infancy~\cite{de2008indicadores}. 

From 2010 onward, productivity rose significantly, despite fewer new research groups being created, but this coincided with the creation of many new graduate programs~\cite{de2020indicadores}, productivity stimulus by the universities due to accreditation requirements, among other circumstances. The productivity curve has not plateaued yet, and we believe the community can further increase its output with over 60 research groups at present in Colombia.

Remarkably, research impact and quality have not been compromised with increased productivity; in fact, they have improved. This aspect is relevant because the number of new professors in the field has not progressed as quickly as the number of graduate students. The fact is that most research groups have reached a maturity level that has allowed them to maximize their resources. Therefore, we believe that the lower share of articles in Q1 journals in recent years is not readily explained by a lower quality of research. Instead, it may be due to traditional optics journals being displaced by new journals in the field.

Concerning the current state of the optics research output from Colombia, it has achieved an adequate performance level--especially impact and quality--fueled by a growing body of researchers and institutions. When compared against other similar countries in Latin America, the Colombian output is well positioned, only superseded by Chile. 

% \rr{Concerning the current state of optics research in Colombia, in agreement with the remarks tendency observed in the big picture, we believe one plausible reason for this results is that the research system in Colombia has probably reached a level of maturity in terms of financial support, policies, equipment, and human resources that has allowed it to position itself higher than many of similar countries in Latin America, and, even though it still has an essential component of international collaboration, if the government keeps and creates the correct policies, in the near future optics research in Colombia will feasibly become more autonomous.}

Nevertheless, the overall impact is partly explained through international collaboration or international leadership, as evidenced by the normalized impact with leadership. However, the scientometric measurement of leadership estimated through the affiliation of the corresponding author is not sufficiently known by the community, which often uses other practices like author-order ranking~\cite{abbasi2021author}. This problem is not exclusive to Colombia, and there is no clear worldwide consensus on the matter~\cite{fernandes2020alphabetic,weber2018effects,kosmulski2012order,mattsson2011correspondence}.  Still, further efforts should be made to improve the impact of research led by Colombian researchers.

% \rr{The reduction in leadership for the case of the research of the highest quality given by the Excellence10wL indicator might have to readings: one is that the international collaboration means that the research, in this case, is led by the international partner. However, a second interpretation could be that there have not been clear policies from the government and institutions to clarify the meaning, and the impact, of appearing in a specific position in the list of authors or as correspondence author. Here is essential to remark that this is not just a problem in the Colombian SNCTI; worldwide there is no consensus on the subject~\cite{fernandes2020alphabetic,weber2018effects,kosmulski2012order,mattsson2011correspondence}.}

The fact that Colombian researchers in optics and related areas are mostly still publishing in non-open-access journals has likely multiple causes. The first is that traditional optics journals are mostly not open access. However, this fact is rapidly changing, with new reputable open-access journals entering the scene and many existing journals transitioning to open-access models. The second apparent cause is the limited funding in the country with no existing plans for open access initiatives like the Plan-S in Europe.

% \rr{Finally, the fact that research in optics and related areas is still published mainly in journals that are not open access can be explained because most traditional journals are not open access or have the option for publications without charges. Nevertheless, this tendency seems to be rapidly changing with the maturity of some of the open-access journals with high impact factors, and the migration of the traditional journals to the open-access system.}

Undoubtedly, each institution's output is directly related to the number of research groups it has, as seen in Fig.~\ref{fig:gruposLoc} and table~\ref{tab:ampo_institutions}. However, this apparent advantage for larger institutions may also play a role against impact, as the number of citations per document suggests. Increased pressure for publication in a research funding scarcity environment may promote quantity over quality, as Fig.~\ref{fig:int-col-vs-ni} suggests. Moreover, accreditation requirements for accessing public funds may have contributed to a notable increase in high-quality publications in small-to-medium size universities that can more easily focus their research efforts than large public institutions.  

%\subsection{Institution-level analysis}
% \rr{It is clear that the number of articles produced by each institution is directly correlated to the number of research groups on each institution, as can be seen by analyzing figure \ref{fig:gruposLoc} and table \ref{tab:ampo_institutions}, basically because there is a bigger workforce for the production of articles. Nevertheless, this also can play against quality, as the number of citations per document suggests, probably because the pressure to generate new ideas is high, resources have to be stretched between more projects and investigators, and a tendency for quantity rather than quality could be built, as figure \ref{fig:int-col-vs-ni} is suggesting.} \rr{Nonetheless, there are two other possible reasons for this kind of behavior: government policies have encouraged small universities and medium private universities to boost the generation of high-quality research to obtain financial support from the government institutions. Alternatively, prominent public universities have the mission to support all kinds of knowledge even if the research associated with it is not as productive in terms of articles production as certain areas of knowledge.}

\section{Conclusions}

The scientometric analysis allowed us to identify research strengths in the Colombian optical community beyond what can be inferred from conventional quantity-centered metrics. Moreover, we showed that optics research in Colombia is an established research area with many active research groups from different institutions spread throughout the country. The research output is primarily concentrated in high-impact journals (Q1 and Q2), achieving a normalized citation impact close to the normalized impact of the world. This impact, however, is mainly driven by international or joint international and national collaboration. Furthermore, the research impact from Colombia is ranked second in Latin America, only superseded by Chile. Finally, the citations from international patents show an opportunity for technological impact. 

% \rr{The scientometric analysis has allowed the identification of research strengths beyond what can be inferred from conventional metrics centered on quantity. Optics research in Colombia has been solidly established, with many active research groups from different institutions. The normalized optic's research impact in Colombia is close to the worldwide impact (0.95), and it is second in Latin America: under Chile (1.33) and over Brasil (0.86), Argentina (0.8), and México (0.77).  However, the relative productivity numbers can still be improved. International collaboration alone and joint international and national collaboration generated high impact research, usually over the average world impact. Most of the production is concentrated in high-impact journals (Q1 and Q2), and the primary communication language is English. There is a clear opportunity to generate a technological impact through international patents.}

\section*{Acknowledgments} %\todo{¿Agradecimientos de Atilio y Andrés?}
E. Rueda thanks Universidad de Antioquia U de A Estrategia de Sostenibilidad “ÓPTICA Y FOTÓNICA-Sostenibilidad 2020-2021”. 
% The authors did not receive specific funding for this work.

\begin{table}[ht]
\centering
\begin{tabular}{@{}lllllllllll@{}}
\toprule
Indicator                   & 2003 & 2004 & 2005 & 2006 & 2007 & 2008 & 2009 & 2010 & 2011 & 2012 \\ \midrule
Output                      & 35   & 33   & 29   & 54   & 35   & 138  & 32   & 61   & 63   & 76   \\
Cites                       & 433  & 571  & 407  & 1449 & 303  & 1220 & 355  & 905  & 778  & 693  \\
Cites per document          & 12.4 & 17.3 & 14.0 & 26.8 & 8.7  & 8.8  & 11.1 & 14.8 & 12.4 & 9.1  \\
\% Cited documents          & 91.4 & 97.0 & 82.8 & 77.8 & 57.1 & 81.2 & 87.5 & 78.7 & 79.4 & 82.9 \\
\% International collab.    & 68.6 & 78.8 & 55.2 & 61.1 & 42.9 & 50.7 & 65.6 & 49.2 & 57.1 & 57.9 \\
\% Int. \& National collab. & 5.7  & 18.2 & 10.3 & 9.3  & 2.9  & 10.9 & 9.4  & 3.3  & 12.7 & 10.5 \\
\% National collab.         & 14.3 & 6.1  & 13.8 & 7.4  & 17.1 & 12.3 & 3.1  & 14.8 & 9.5  & 10.5 \\
\% Without collab.          & 17.1 & 15.2 & 31.0 & 31.5 & 40.0 & 37.0 & 31.3 & 36.1 & 33.3 & 31.6 \\
Output in Q1                & 11   & 16   & 10   & 20   & 12   & 18   & 11   & 10   & 15   & 19   \\
\% Output in Q1             & 31.4 & 48.5 & 34.5 & 37.0 & 34.3 & 13.0 & 34.4 & 16.4 & 23.8 & 25.0 \\
Leadership                  & 28   & 21   & 24   & 39   & 31   & 121  & 24   & 53   & 45   & 61   \\
\% Leadership               & 80.0 & 63.6 & 82.8 & 72.2 & 88.6 & 87.7 & 75.0 & 86.9 & 71.4 & 80.3 \\
Normalized Impact           & 0.6  & 0.8  & 0.7  & 1.5  & 0.4  & 0.6  & 0.6  & 0.7  & 0.8  & 0.6  \\
Normalized Impact wL        & 0.7  & 0.5  & 0.7  & 1.6  & 0.3  & 0.7  & 0.6  & 0.5  & 0.8  & 0.6  \\
Excellence10                & 2    & 3    & 3    & 8    & 1    & 7    & 2    & 7    & 4    & 5    \\
\% Excellence10             & 5.7  & 9.1  & 10.3 & 14.8 & 2.9  & 5.1  & 6.3  & 11.5 & 6.4  & 6.6  \\
Excellence10 wL             & 1    & 2    & 2    & 3    & 1    & 6    & 1    & 4    & 3    & 5    \\
\% Excellence10 wL          & 2.9  & 6.1  & 6.9  & 5.6  & 2.9  & 4.4  & 3.1  & 6.6  & 4.8  & 6.6  \\
Open Access                 & 2    & 2    & 8    & 4    & 0    & 12   & 3    & 3    & 9    & 14   \\
\% Open Access              & 5.7  & 6.1  & 27.6 & 7.4  & 0.0  & 8.7  & 9.4  & 4.9  & 14.3 & 18.4 \\ \bottomrule
\end{tabular}
\caption{General productivity and scientometric indicators for the Atomic and Molecular Physics, and Optics (AMPO) category for Colombia between 2003-2020.}
\label{tab:01a}
\end{table}

\addtocounter{table}{-1} %to continue the table

\begin{table}[h!]
\centering
\begin{tabular}{@{}lllllllll@{}}
\toprule
Indicator                   & 2013 & 2014 & 2015 & 2016 & 2017 & 2018 & 2019 & 2020 \\ \midrule
Output                      & 93   & 125  & 139  & 114  & 159  & 174  & 196  & 202  \\
Cites                       & 1407 & 1670 & 955  & 1734 & 1628 & 1227 & 700  & 182  \\
Cites per document          & 15.1 & 13.4 & 6.9  & 15.2 & 10.2 & 7.1  & 3.6  & 0.9  \\
\% Cited documents          & 83.9 & 72.0 & 73.4 & 91.2 & 80.5 & 82.8 & 63.8 & 39.1 \\
\% International collab.    & 63.4 & 65.6 & 54.7 & 64.9 & 67.9 & 62.6 & 51.0 & 61.9 \\
\% Int. \& National collab. & 8.6  & 6.4  & 7.2  & 7.9  & 12.0 & 16.7 & 8.7  & 11.9 \\
\% National collab.         & 6.5  & 8.0  & 13.0 & 11.4 & 12.0 & 17.2 & 19.9 & 19.8 \\
\% Without collab.          & 30.1 & 26.4 & 32.4 & 23.7 & 20.1 & 20.1 & 29.1 & 18.3 \\
Output in Q1                & 30   & 30   & 30   & 23   & 35   & 64   & 49   & 61   \\
\% Output in Q1             & 32.3 & 24.0 & 21.6 & 20.2 & 22.0 & 36.8 & 25.0 & 30.2 \\
Leadership                  & 66   & 91   & 106  & 77   & 111  & 118  & 133  & 142  \\
\% Leadership               & 71.0 & 72.8 & 76.3 & 67.5 & 69.8 & 67.8 & 67.9 & 70.3 \\
Normalized Citation         & 0.89 & 1.46 & 0.71 & 1.22 & 0.95 & 0.90 & 0.89 & 0.90 \\
Normalized Citation wL      & 0.78 & 0.58 & 0.59 & 0.71 & 0.86 & 0.72 & 0.56 & 0.78 \\
Excellence10                & 14   & 15   & 9    & 15   & 19   & 16   & 11   & 24   \\
\% Excellence10             & 15.1 & 12.0 & 6.5  & 13.2 & 12.0 & 9.2  & 5.6  & 11.9 \\
Excellence10 wL             & 9    & 8    & 5    & 5    & 9    & 5    & 6    & 13   \\
\% Excellence10 wL          & 9.7  & 6.4  & 3.6  & 4.4  & 5.7  & 2.9  & 3.1  & 6.4  \\
Open Access                 & 13   & 20   & 17   & 24   & 48   & 53   & 47   & 57   \\
\% Open Access              & 14.0 & 16.0 & 12.2 & 21.1 & 30.2 & 30.5 & 24.0 & 28.2 \\ \bottomrule
\end{tabular}
\caption{(\textit{Continued}) General productivity and scientometric indicators for the Atomic and Molecular Physics, and Optics (AMPO) category for Colombia between 2003-2020.}
\label{tab:01b}
\end{table}

\iffalse
\section{Background methodology}\label{Appe:BG method}
In first place, we used the data available from the Colombian Optical Network to contact the optics research groups directly and ask for information related to members, lines of research, and the group's history. Then, the information delivered by the research groups was complemented with the data available in GrupLac. The next step consisted of searching in GrupLac all the research groups with names associated with optics or related areas such as photonics, spectroscopy, image, and lasers. Finally, using the database of authors obtained with Scopus, we used CvLac and GrupLac to find the research groups that are associated to each author. With this information, we built the groups' background history, location and timeline of establishment.
\fi

%%%%%%%%%% If using BibTeX:
\bibliography{sample}

%%%%%%%%%% If preparing manually:
% \newcommand{\enquote}[1]{``#1''}

% \bibitem{Zhang:14}
% Y.~Zhang, S.~Qiao, L.~Sun, Q.~W. Shi, W.~Huang, L.~Li, and Z.~Yang,
%   \enquote{Photoinduced active terahertz metamaterials with nanostructured
%   vanadium dioxide film deposited by sol-gel method,}
%   {\protect\JournalTitle{Optics Express}} \textbf{22}, 11070--11078 (2014).

% {Optical Society}, \enquote{{OSA Publishing},}
%   \url{http://www.osapublishing.org}.

% \bibitem{FORSTER2007}
% P.~Forster, V.~Ramaswamy, P.~Artaxo, T.~Bernsten, R.~Betts, D.~Fahey,
%   J.~Haywood, J.~Lean, D.~Lowe, G.~Myhre, J.~Nganga, R.~Prinn, G.~Raga,
%   M.~Schulz, and R.~V. Dorland, \enquote{Changes in atmospheric consituents and
%   in radiative forcing,} in \enquote{Climate Change 2007: The Physical Science
%   Basis. Contribution of Working Group 1 to the Fourth assesment report of
%   Intergovernmental Panel on Climate Change,}  S.~Solomon, D.~Qin, M.~Manning,
%   Z.~Chen, M.~Marquis, K.~B. Averyt, M.~Tignor, and H.~L. Miler, eds.
%   (Cambridge University Press, 2007).

% \end{thebibliography}

\section*{Appendix}

In the following table we include the full scientometric indicators for the 2003-2020 period.

\end{document}